\documentclass[5p]{elsarticle}
\usepackage{graphicx,bm,times,miller}
\usepackage{natbib}

\begin{document}

\title{Comparing the Corrosion of Uranium Nitride and Uranium Dioxide Surfaces with H$_2$O$_2$} 

\author[1]{E. Lawrence Bright}
\cortext[mycorrespondingauthor]{Corresponding author.}
\ead{e.lawrencebright@bristol.ac.uk}
\author[1]{S. Rennie}
\author[1]{A. Siberry}
\author[1]{K. Samani}
\author[1]{K. Clarke}
\author[2]{D.T. Goddard}
\author[1]{R. Springell}

\address[1]{School of Physics, University of Bristol, Tyndall Avenue, Bristol BS8 1TL, UK.}
\address[2]{National Nuclear Laboratory, Preston Laboratory, Springfields, Preston, Lancashire PR4 0XJ, UK.}

\begin{abstract}

Uranium mononitride, UN, is considered a potential accident tolerant fuel due to its high uranium density, high thermal conductivity, and high melting point.
Compared with the relatively inert UO$_2$, UN has a high reactivity in water, however, studies have not considered the significant effect of radiation, which is known to cause corrosion of UO$_2$.
This study uses 0.1\,M H$_2$O$_2$ to simulate the effects of water radiolysis in order to compare the radiolytic corrosion rates of UO$_2$, UN, and U$_2$N$_3$ thin films at room temperature. 
X-ray reflectivity was used to investigate the changes in film morphology as a function of H$_2$O$_2$ exposure time, allowing changes in film thickness and roughness to be observed on the \AA{}ngstrom length-scale.
Results showed significant differences between UO$_2$, UN, and U$_2$N$_3$, with corrosion rates of  0.083(3), 0.020(4), and 0.47(8)\,\AA{}/s, respectively, showing that UN corrodes more slowly than UO$_2$ in 0.1\,M H$_2$O$_2$.

\end{abstract}

\date{\today}

\maketitle

\section{Introduction}

Accident tolerant fuels (ATFs) are a key concept in the drive towards increased passive safety in the nuclear industry \cite{Zinkle2014}. 
The 2011 Fukushima Daiichi accident highlighted the thermal limitations of the current UO$_2$-Zr fuel-cladding system and as such, significant effort is being invested in researching accident tolerant alternatives.
In order to offset the manufacturing and licensing costs associated with introducing a new fuel or cladding material, a secondary aim of ATFs is to improve fuel economy by potentially allowing for increased burn-up and cycle lengths. 

Of the alternative fuels that have been explored, UN has been highlighted as a prime candidate, with higher uranium density, thermal conductivity, and similarly high melting point in comparison with UO$_2$ \cite{Evans1963,Ross1988}. 
Despite this, UN has two considerable disadvantages: (1) its high reactivity in water and air above 200\,$^\circ$C; (2) a large N$^{14}$ cross-section that results in a substantial amount of C$^{14}$ production \cite{Sugihara1969,Sunder1998,Jolkkonen2017,Paljevic1975}. 
While the latter can be resolved by enriching UN with N$^{15}$, the high reactivity with water is considered a significant road block to UN being selected as a nuclear fuel in water-moderated fission reactors. 
Before UN is to be ruled out as a potential ATF, this issue should be explored further, such that the interactions that take place at the fuel / water interface are well-understood, allowing improved prediction or even mitigation of this reaction. 
Work has been conducted to characterise the corrosion behavior of UN in water, where its performance has been poor; however, there is no research addressing the effect of high radiation fields \cite{Sugihara1969,Sunder1998,Dell1967}.
Given that the fuel is most likely to be exposed to water either as a result of a fuel rod failure during reactor operation, or during storage as spent nuclear fuel, strong radiation fields will be present and must be considered. 

In the presence of a strong radiation field, water is quickly radiolysed, giving rise to highly oxidising species, including H$_2$O$_2$, OH$^{\bullet}$, H$^{\bullet{}}$, HO$_2^{\bullet}$, and e$_{eq}^{-}$ \cite{LeCaer2011,Spinks1990}. 
In the case of UO$_2$, which is insoluble in neutral water, interactions with these species leads to the formation of the readily soluble U$^{6+}$ ion, causing dissolution of the fuel matrix \cite{Shoesmith2007}.   
As such, it is necessary for the behavior of UN to be examined within this radiolytic environment and compared to UO$_2$, in order to assess its reactivity in an accident scenario and, thus assess its viability as an ATF. 
It is our contention that these conditions should be the primary concern, and not simply the water / fuel interaction, as a comparison of fuel surfaces in these scenarios could lead to drastically different results, as has shown to be the case with UO$_2$ \cite{Bailey1985,Shoesmith1992}. 

Until this work, there have been no studies to investigate the interaction of UN with radiolytically produced, oxidising species. 
Such studies are regularly performed on UO$_2$, either using a radiation source to induce water radiolysis or through chemically simulating the radiolytic products \cite{Bailey1985,Shoesmith1992,Sunder1997,Shoesmith2000,Springell2015}.
The latter approach is most commonly achieved using hydrogen peroxide, H$_2$O$_2$, a long-lived, highly oxidising species produced during water radiolysis. 

In this study, we monitor the change in surface morphology of UN, U$_2$N$_3$, and UO$_2$ thin films as a function of H$_2$O$_2$ exposure, using x-ray reflectivity (XRR). 
Thin film samples are optimal for studying radiolytic dissolution, as these idealised surfaces enable the changes at the film / water interface to be observed on the \AA{}ngstrom length-scale. 
Using this approach, the corrosion behavior of UN and UO$_2$ nanocrystalline thin films has been compared, where UO$_2$ provides the benchmark for fuel behavior. 
As the oxidation of UN is known to progress through the formation of a U$_2$N$_3$ interlayer, a nanocystalline U$_2$N$_3$ film has also been investigated.
This experiment will therefore give the first insight into the potential corrosion rate and mechanism of UN in an accident scenario.

\section {Experiment}

Thin film samples were grown using reactive DC magnetron sputtering of a uranium target in a partial pressure of nitrogen or oxygen, with 5.5\,N argon used as the main sputtering gas at a pressure of 7x10$^{-3}$\,mbar.
UO$_2$ films were grown in 2x10$^{-5}$\,mbar O$_2$, UN films were grown in 2x10$^{-5}$\,mbar N$_2$, and U$_2$N$_3$ films were grown in 9x10$^{-4}$\,mbar N$_2$.
Films were deposited at room temperature at a thickness of roughly 600\,\AA{} on highly polished corning glass substrates, supplied by MTI Corporation.
For consistency across the corrosion  measurements, samples of the same material were grown simultaneously and divided into 5$\times$5\,mm individual samples.

In order to measure grain size across all materials, thicker films of roughly 1000\,\AA{} of UO$_2$, UN, and U$_2$N$_3$ were grown under the same conditions and, in the case of the UN and U$_2$N$_3$ films, capped with 40\,\AA{} of niobium to prevent oxidation.
X-ray diffraction (XRD) was performed on these samples, using a Philips X'Pert diffractometer with a Cu-K$\alpha$ source, to identify phases present in each sample and calculate crystallite size using the Scherrer equation \cite{Patterson1939}.

\begin{figure} \centering \includegraphics[width=0.9\linewidth]{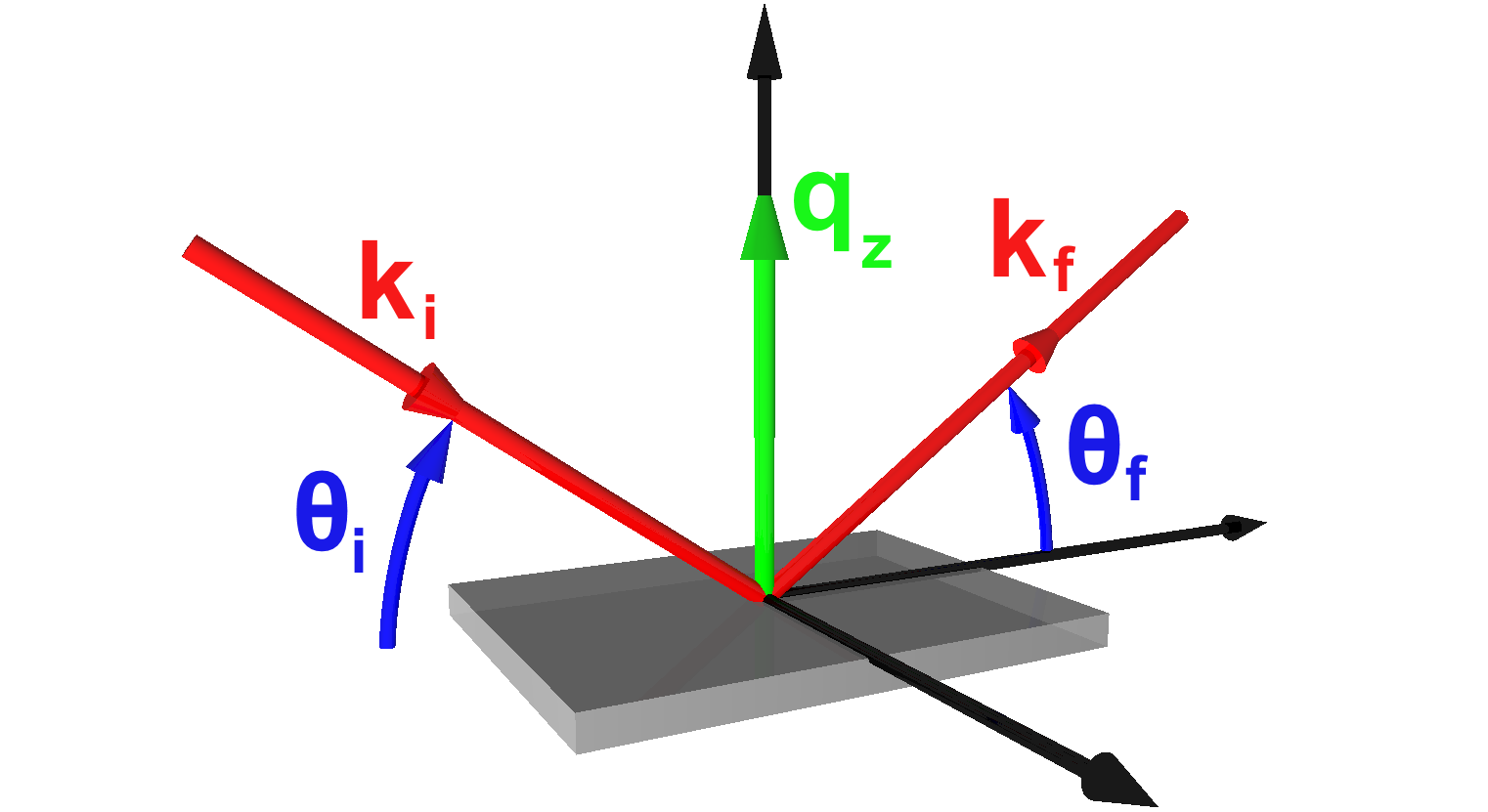} 
\caption{Diagram showing XRR set-up, with the incident and exit wavevectors (\textbf{k$_i$} and \textbf{k$_f$}), angle of incidence and exit ($\theta_i$ and $\theta_f$), and the wavevector momentum transfer (\textbf{q$_z$}) labeled.\label{fig:xrr}} 
\end{figure}

XRR was used as a non-destructive method of probing the morphology of the layers in the thin film samples before and after corrosion in H$_2$O$_2$.
This technique probes the electron density, or scattering length density, of a material perpendicular to its surface, by measuring intensity as a function of \textbf{q$_z$}, the wavevector momentum transfer.
This wavevector is the difference between the incident and exit wavevectors, \textbf{k$_i$} and \textbf{k$_f$}, respectively, as displayed in Fig. \ref{fig:xrr}.
In order to measure only in the specular direction, the direction perpendicular to the surface, the angles of incidence and exit, $\theta_i$ and $\theta_f$, are kept equal.

At very low \textbf{q$_z$}, x-rays will be reflected at the surface of the sample as the refractive index of air is higher than that of the sample.
As \textbf{q$_z$} increases, at an angle known as the critical angle, the x-rays will partially penetrate the sample and the reflected intensity will decrease significantly.  
This critical angle is dependent on the refractive index and therefore electron density of the sample \cite{Tolan1995}.
Beyond the critical angle, x-rays that penetrate the surface reflect and refract at each interface, with the reflected rays interfering, giving rise to Keissig fringes in the reflectivity profile.
The observed separation between fringes is inversely proportional to the distance between interfaces \cite{Parratt1954}. 
For a non-ideal film, roughening between interfaces causes diffuse scattering, reducing the specular reflected intensity, decreasing the resolution of these interference fringes \cite{Nevot1980}.  

To extract this morphological information from XRR, data was modeled using GenX, a software package that utilises a differential evolution fitting algorithm to simulate and fit reflectivity data using the Parratt recursion method \cite{Bjorck2007,Parratt1954}.
The scattering length density (proportional to electron density) profile is modeled as a function of depth through the sample.
This scattering length density (SLD) plot is described by a series of layers comprising the substrate, film, and oxidised surface, with each layer defined by a fixed electron density. 
While the substrate is defined as being infinitely thick, the thickness of the film and oxidised surface are allowed to vary, along with the roughness of each interface.
This roughness is modeled as a gradient between electron density at each boundary, and described with a Gaussian distribution, with the roughness value obtained being the root mean square roughness in \AA{} \cite{Nevot1980}.

For consistency across data sets when modeling and fitting data, all instrument parameters were kept constant, with only film and oxidised layer thickness, interface roughness, and top layer stoichiometry allowed to vary between fits.
UN and U$_2$N$_3$ films were modeled with an oxidised surface layer consisting of UO$_2$, the thickness of which was allowed to vary \cite{Dell1967}.
There is evidence in the literature of a U$_2$N$_3$ interlayer forming between UN and UO$_2$ during the oxidation of UN (and this is the reason for the inclusion of U$_2$N$_3$ in this study), however, the very similar electron densities of UN and U$_2$N$_3$ make them indistinguishable with XRR \cite{Dell1967,Rao1991}. 
For this reason, a U$_2$N$_3$ interlayer was not included in the model of UN.
UO$_2$ films were modeled as uniform, stoichiometric UO$_2$, while the surface of UO$_2$ films were modeled as hyperstoichiometric UO$_{2+x}$, where electron density was allowed to vary from the stoichiometric value.
This hyperstoichiometric UO$_{2+x}$ surface layer was also included when modeling the oxidised surfaces of UN and U$_2$N$_3$ films.  

Error values for these fitted parameters were calculated as a 5\,\% increase in the figure of merit, calculated using Eq. \ref{fom}, where $i$ are all Q values, $D$ is normalised reflectivity data, and $M$ is modeled reflectivity. 
This method of calculating errors indicates how much altering a value in the model worsens the fit of the model.

\begin{equation} \label{fom}
FOM = \frac{\sum_{i}^{}\left | \log_{10}D_i - \log_{10}M_i \right |}{\sum_{i}^{} \log_{10} D_i }
\end{equation}

To investigate the effects of radiation on corrosion rates, 0.1\,M hydrogen peroxide was used to simulate the radiolytic products of water. 
While this concentration is significantly higher than that which can be expected at the surface of irradiated fuel in water, it has been selected to reproduce the high corrosion rates seen when this scenario is simulated with synchrotron radiation and that which would be expected in an accident scenario \cite{Springell2015}.
Additionally, using a higher H$_2$O$_2$ concentration allows experiments to be performed on a shorter timescale, meaning the change in concentration due to the decomposition of H$_2$O$_2$ does not introduce significant errors into the experiment. 
Corrosion experiments were therefore performed by submerging samples in 0.1\,M H$_2$O$_2$ for either 50\,s, 250\,s, 1250\,s, 6000\,s, or 4 exposures of 1250\,s at room temperature, with XRR measured before and after exposure.
The volume of solution used was chosen to be large enough such that the saturation limit of the solution would not be reached after complete corrosion.

\section{Results}

\begin{figure} \includegraphics[width=1\linewidth]{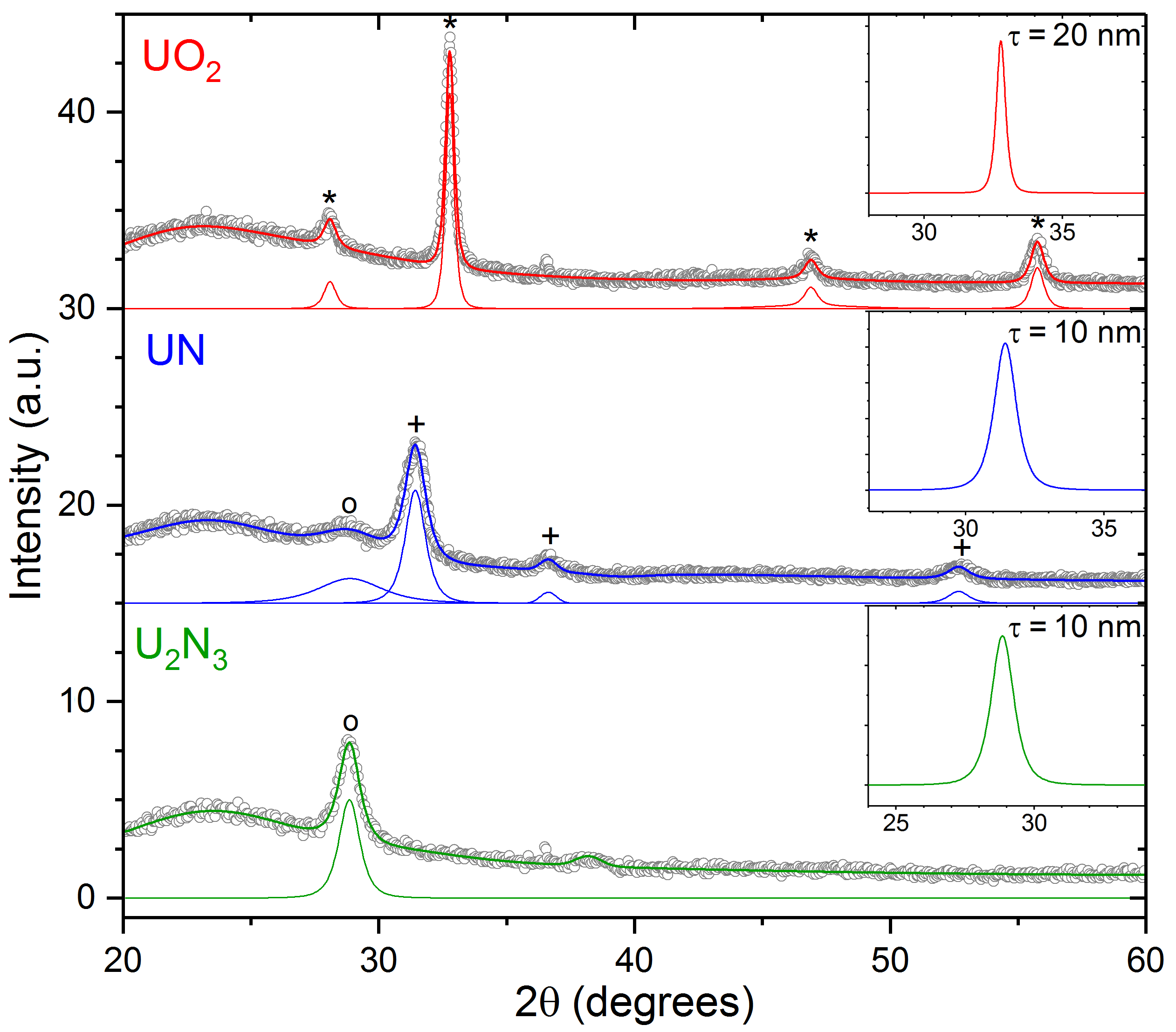}\caption{Fitted XRD scan of UO$_2$ and capped UN and U$_2$N$_3$ thin films. Peaks arising from UO$_2$ are labeled with an asterisk, UN labeled with a cross and U$_2$N$_3$ labeled with a circle. \label{fig:xrd}} \end{figure} 

Structural characterisation was carried out on the UO$_2$ and capped UN and U$_2$N$_3$ thin films, using XRD, as shown in Fig \ref{fig:xrd}.
All scans show a broad amorphous background, with a peak at 23\,$^\circ$, from the corning glass substrates. 
The film shown to be UO$_2$ is single phase, with peaks of FWHM of 0.39\,$^\circ$ in 2$\theta$, when fitted with a Gaussian, as shown inset in Fig. \ref{fig:xrd}.
This corresponds to a crystallite size, $\tau$, of 20$\pm$2\,nm when the Scherrer equation is used with a shape factor of 0.9.
The UN film shows some presence of U$_2$N$_3$ contamination, labeled with a circle, while the U$_2$N$_3$ film was found to be single phase.
The small peak observed at 38\,$^\circ$ for the U$_2$N$_3$ film is attributed to the aluminium sample stage.
Fitting of the highest intensity peaks of UN and U$_2$N$_3$ gave a FWHM of 0.93\,$^\circ$ and 0.96\,$^\circ$, respectively, as shown inset in Fig. \ref{fig:xrd}, found to correspond to a crystallite size, $\tau$, of 10$\pm$1\,nm for both UN and U$_2$N$_3$.
The relative intensities of the Bragg peaks show some preferred orientation in the \hkl[111] direction for UN and U$_2$N$_3$ and in the \hkl[001] direction for UO$_2$.
Peak positions for the films show UO$_2$ to be stoichiometric and UN and U$_2$N$_3$ to be slightly hyperstoichiometric, within the errors of the measurements.

\begin{figure*} \centering \includegraphics[width=0.95\linewidth]{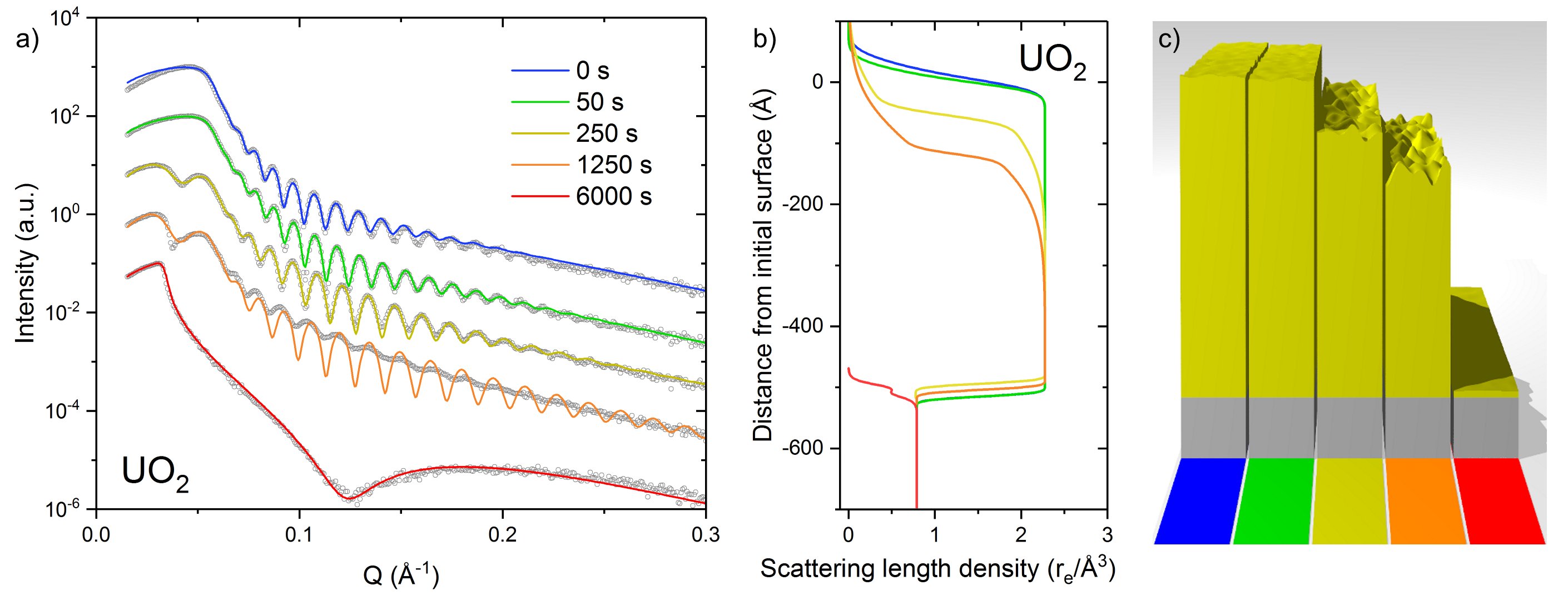} 
\caption{Graphs showing a) fitted XRR of UO$_2$ samples after 0\,s, 50\,s, 250\,s, 1250\,s, and 6000\,s of exposure to H$_2$O$_2$, with data represented by open grey circles and fits by solid lines, b) SLD as a function of depth from the XRR model, and c) schematic illustrations of UO$_2$ thin film samples after each exposure time. \label{fig:uo2}} \end{figure*}

\begin{figure*} \centering \includegraphics[width=0.95\linewidth]{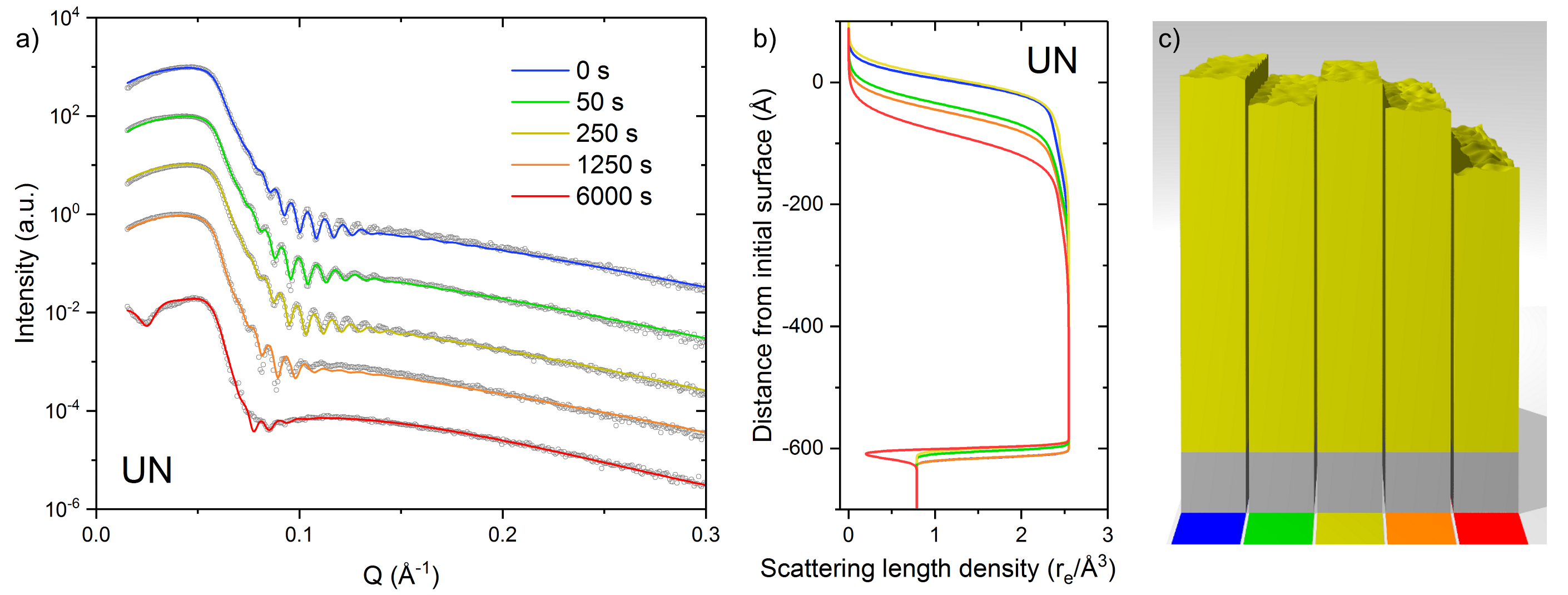} 
\caption{Graphs showing a) fitted XRR of UN samples after 0\,s, 50\,s, 250\,s, 1250\,s, and 6000\,s of exposure to H$_2$O$_2$, with data represented by open grey circles and fits by solid lines, b) SLD as a function of depth from the XRR model, and c) schematic illustrations of UN thin film samples after each exposure time. \label{fig:un}} \end{figure*}

\begin{figure*} \centering \includegraphics[width=0.95\linewidth]{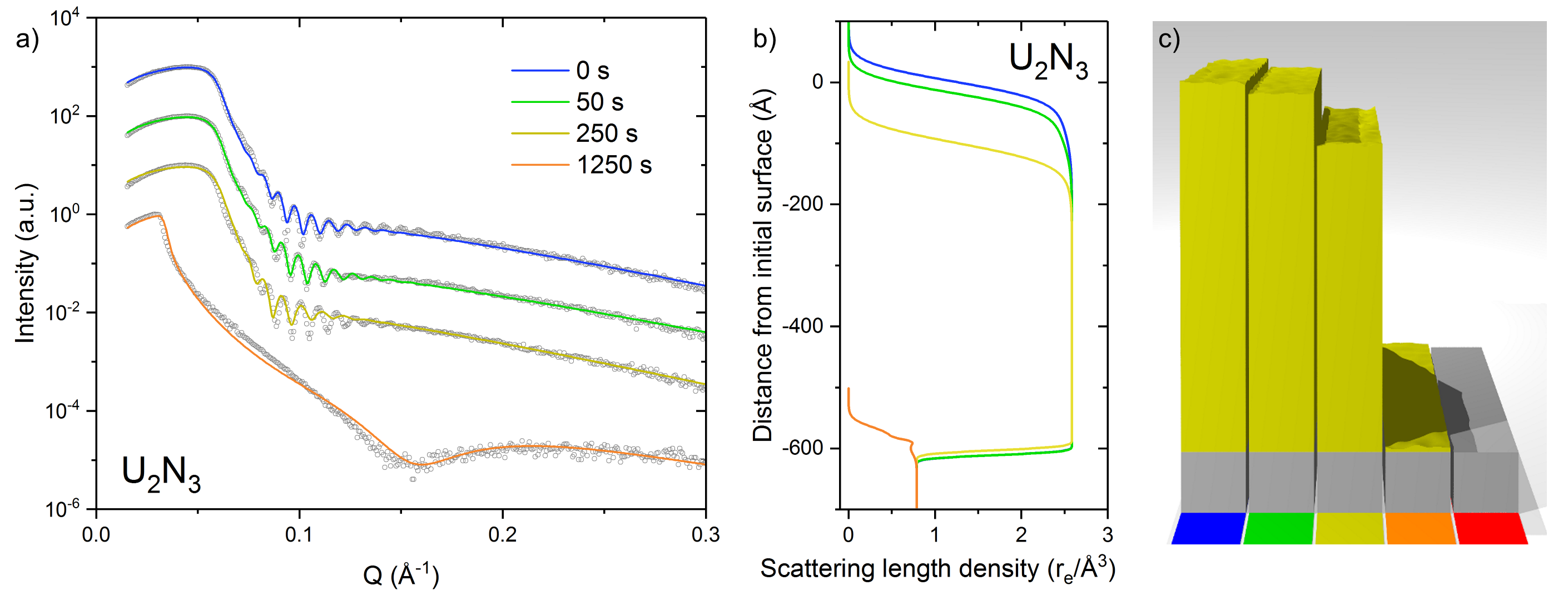} 
\caption{Graphs showing a) fitted XRR of U$_2$N$_3$ samples after 0\,s, 50\,s, 250\,s, and 1250\,s of exposure to H$_2$O$_2$, with data represented by open grey circles and fits by solid lines, b) SLD as a function of depth from the XRR model, and c) schematic illustrations of U$_2$N$_3$ thin film samples after each exposure time. \label{fig:u2n3}} \end{figure*}

Fig. \ref{fig:uo2} shows the effect of exposure to 0.1\,M H$_2$O$_2$ for varying time periods on thin film samples of polycrystalline UO$_2$.
Section a) shows plots of the XRR data taken before and after exposures of varying times, with the modeled fits shown as solid lines, and corresponding SLD plots of the models shown in b).

The increase in fringe width in the XRR of UO$_2$ samples shows the decrease in sample thickness with increasing exposure time to H$_2$O$_2$, as can be seen in the SLD plots.
This trend continues as a function of time until almost all the film has been corroded. 
After 6000\,s, the critical angle has moved to Q = 0.03\,\AA$^{-1}$, corresponding to the density of the substrate only.

From the SLD plots it can be seen that the roughness of the UO$_2$ thin film samples initially decreases after 50 s exposure to H$_2$O$_2$, before increasing significantly with exposure time.
This is visible in the XRR data from the decrease in fringe depth and the appearance of a second critical angle at Q = 0.03 \AA$^{-1}$, corresponding to the density of the substrate, after 250\,s. 
This gives evidence that there are areas of the sample where the substrate is exposed and therefore where the film has completely corroded.
After 1250\,s, the relative intensity of the critical angle at Q = 0.03\,\AA$^{-1}$, is greater than at 250\,s, suggesting that there is a larger area of the substrate exposed.
At this time point, there is also a large discrepancy in the XRR fringe depth between the data and fitted model, indicating that the model is not accurate, and should be rougher at the surface.

The XRR measurements and modeled fits of polycrystalline UN thin film samples exposed to H$_2$O$_2$ for varying times are shown in section a) of Fig. \ref{fig:un}, with SLD plots in section b).
Measurements of the UN thin film samples show a small increase in fringe width after 50\,s of exposure, corresponding to a total thickness decrease of 30\,\AA{}, shown clearly in the SLD plot.
The UN thin film sample exposed to H$_2$O$_2$ for 250\,s shows very little change, with total film thickness remaining almost the same.
Similar to that exposed for 50\,s, the sample exposed for 1250\,s showed a decrease in thickness of around 30\,\AA{}.
This trend continues, with the sample exposed for 6000\,s showing a decrease in thickness of 150\,\AA{}.

While the position of the surface and oxidised layer shown in the SLD plot changes for different exposure times, the shape of the curve does not change significantly for exposure times up to 1250\,s.
This shows that the oxidised layer thickness and surface roughness do not change significantly as a function of exposure time.
However, some small changes in roughness are seen in the SLD plot of the sample exposed for 6000\,s.
This is supported by the presence of a second critical angle, suggesting that there are areas of the film that have been completely corroded.
Additionally, the depth of the fringes is appreciably smaller in this XRR measurement, showing increased roughness.

Section a) of Fig. \ref{fig:u2n3} shows the XRR of polycrystalline U$_2$N$_3$ thin films exposed to H$_2$O$_2$ for varying times and modeled fits, with the SLD plots of these models shown in section b).
The decreasing film thickness as a function of exposure time is clearly shown in the SLD plot and is evident in the increasing fringe thickness visible in the XRR data, 
with the film exposed to H$_2$O$_2$ for 1250\,s being almost completely corroded.
This can be seen from the critical angle of the XRR corresponding to the density of the substrate and broad fringe showing the low thickness of any remaining film.

From the constant fringe depth in the XRR of U$_2$N$_3$ after exposures of 50\,s and 250\,s, it can be seen that the roughness of the film surface stays constant, as shown in the similar shape in the SLD plots.
In addition, there is no appearance of a second critical angle until the film is completely corroded after 1250\,s, showing that corrosion occurs uniformly over the U$_2$N$_3$ sample.

For clarity, illustrations of samples after exposure to H$_2$O$_2$ for varying times have been produced and are shown in section c) of  Fig. \ref{fig:uo2}, \ref{fig:un}, and \ref{fig:u2n3}, derived from calculated values of thickness and roughness, where the relative changes are to scale. 

\begin{figure} \includegraphics[width=1\linewidth]{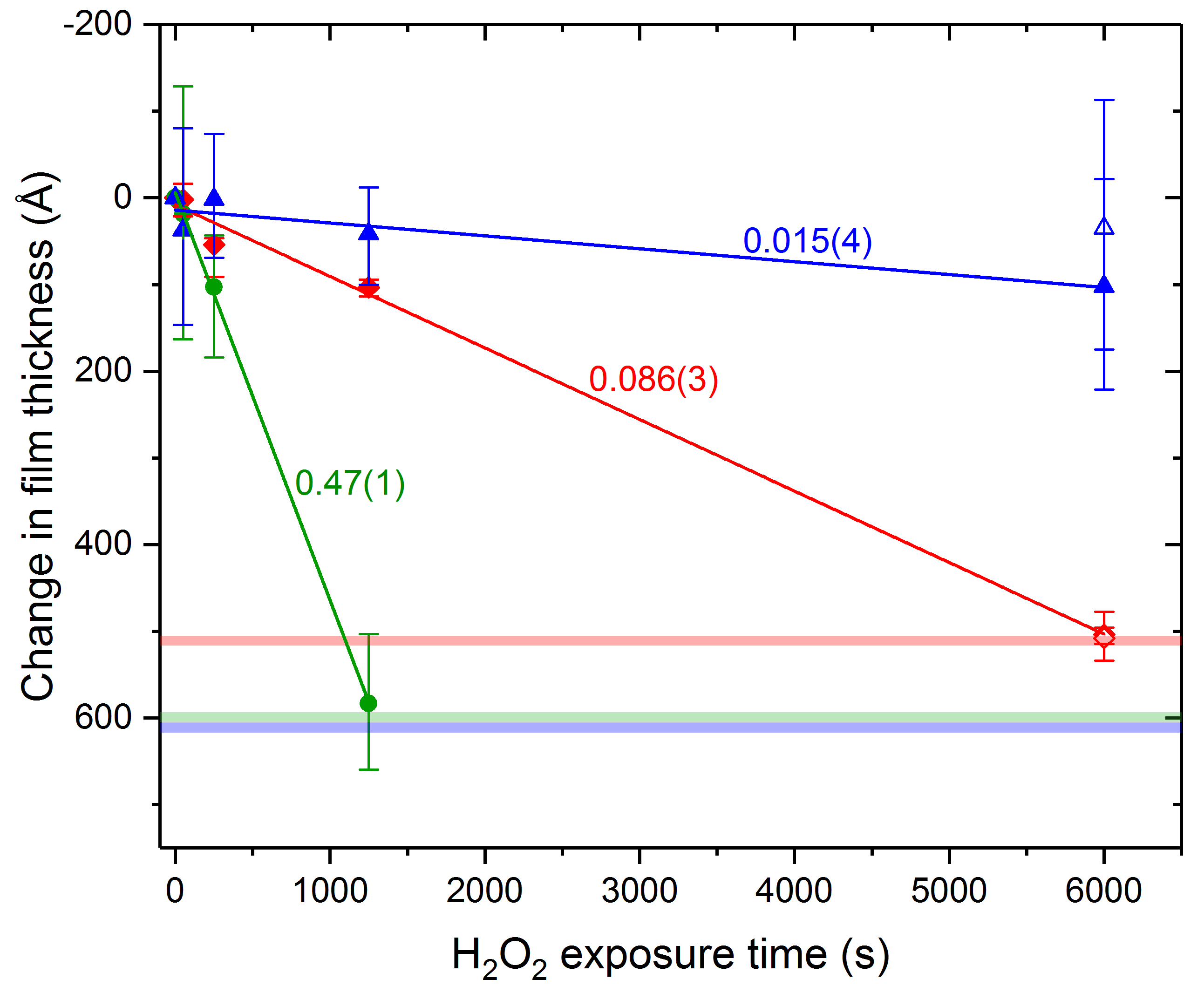}
\caption{Graph showing change in sample thickness of UO$_2$ (red diamond), UN (blue triangle), and U$_2$N$_3$ (green circle) as a function of exposure time to H$_2$O$_2$. Closed points show single exposures and open points show the cumulative time after 4 exposures of 1250\,s. The pale lines denote the initial total sample thickness and the solid line shows a linear fit to the data, labelled with the gradient.  \label{fig:tvt}} \end{figure} 

Fig. \ref{fig:tvt} shows a comparison of total thickness changes as a function of exposure time to H$_2$O$_2$ between polycrystalline UO$_2$, UN, and U$_2$N$_3$ thin film samples.
As initial film thickness was not identical between materials, the negative total initial thickness for each material (and therefore maximum thickness that can be lost by samples of each material) is shown by the light horizontal lines in red for UO$_2$, blue for UN, and green for U$_2$N$_3$.
Closed points show the result of a single exposure and are fitted with straight solid lines, with the gradient labeled, while open points show the cumulative time after 4 exposures.
The large difference in gradients between the linear fits shows the difference in corrosion rates between UO$_2$, UN, and U$_2$N$_3$, with UN being the slowest and U$_2$N$_3$ the fastest to corrode.

\begin{figure}[h!] \includegraphics[width=1\linewidth]{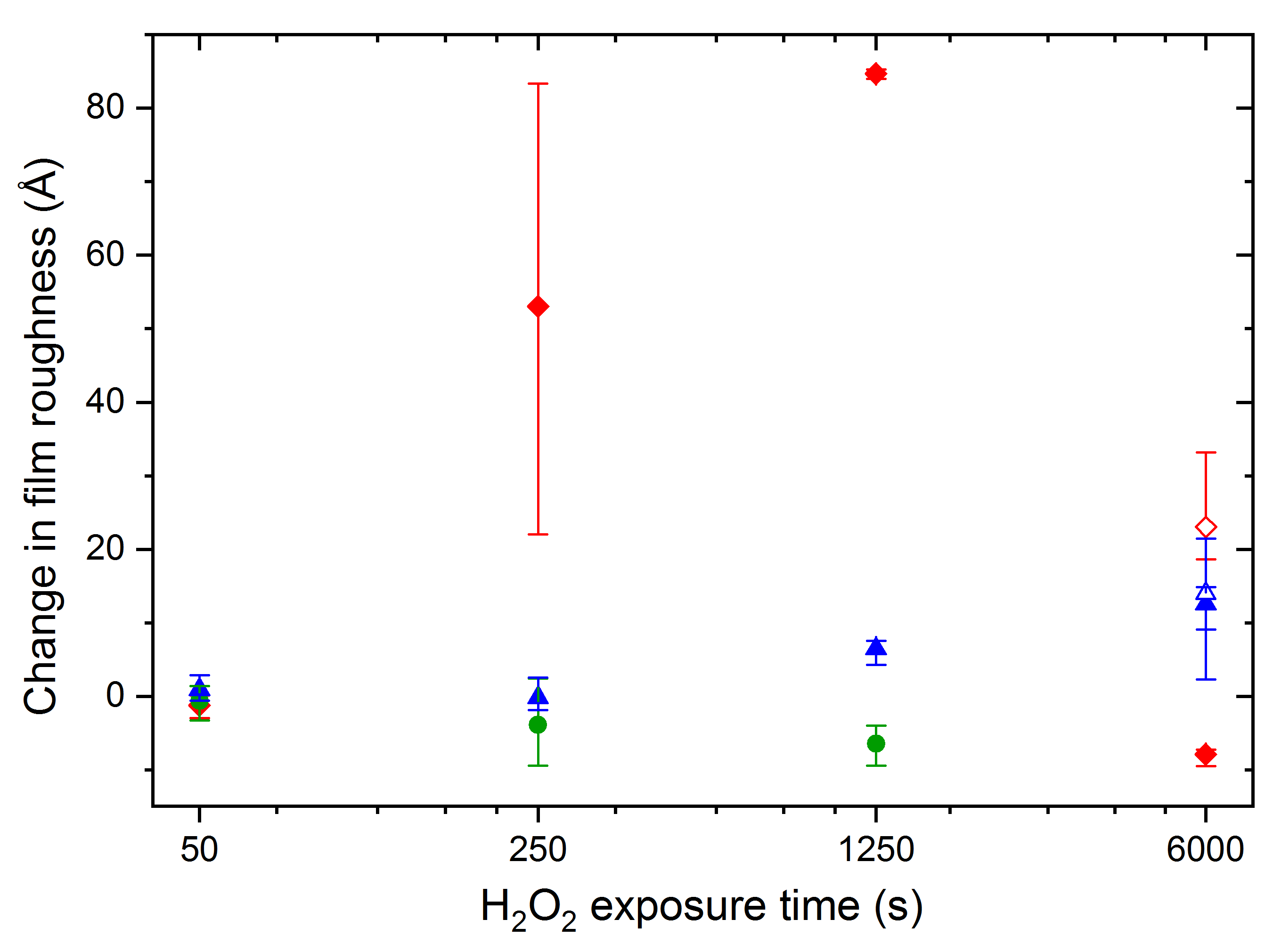}
\caption{Graph showing change in sample roughness of UO$_2$ (red diamond), UN (blue triangle), and U$_2$N$_3$ (green circle) as a function of exposure time to H$_2$O$_2$. Closed points show single exposures and open points show the cumulative time after 4 exposures of 1250\,s.\label{fig:rvt}} \end{figure} 

The change in surface roughness of the UO$_2$, UN, and U$_2$N$_3$ samples as a function of H$_2$O$_2$ exposure time is displayed in Fig. \ref{fig:rvt}.
Roughness of UO$_2$ samples increases significantly with exposure time, with the exception being the roughness change after 6000\,s, when the film has almost completely corroded.
There is a difference of 30\,\AA{} RMS roughness between 4$\times$1250\,s and 6000\,s, greater than the errors on both data points.
However, with the UO$_2$ film being almost completely corroded for both time points, the measured roughness is decreased by the low roughness of the substrate. 
UN shows a small increase in roughness as a function of H$_2$O$_2$ exposure time, with the roughness change after 6000\,s and 4x1250\,s being the same.
U$_2$N$_3$ shows no change in roughness with H$_2$O$_2$ exposure, except for a small decrease at 1250\,s, when the film has almost completely corroded and the roughness measured is effectively that of the glass substrate.

\begin{figure}[h!] \includegraphics[width=1\linewidth]{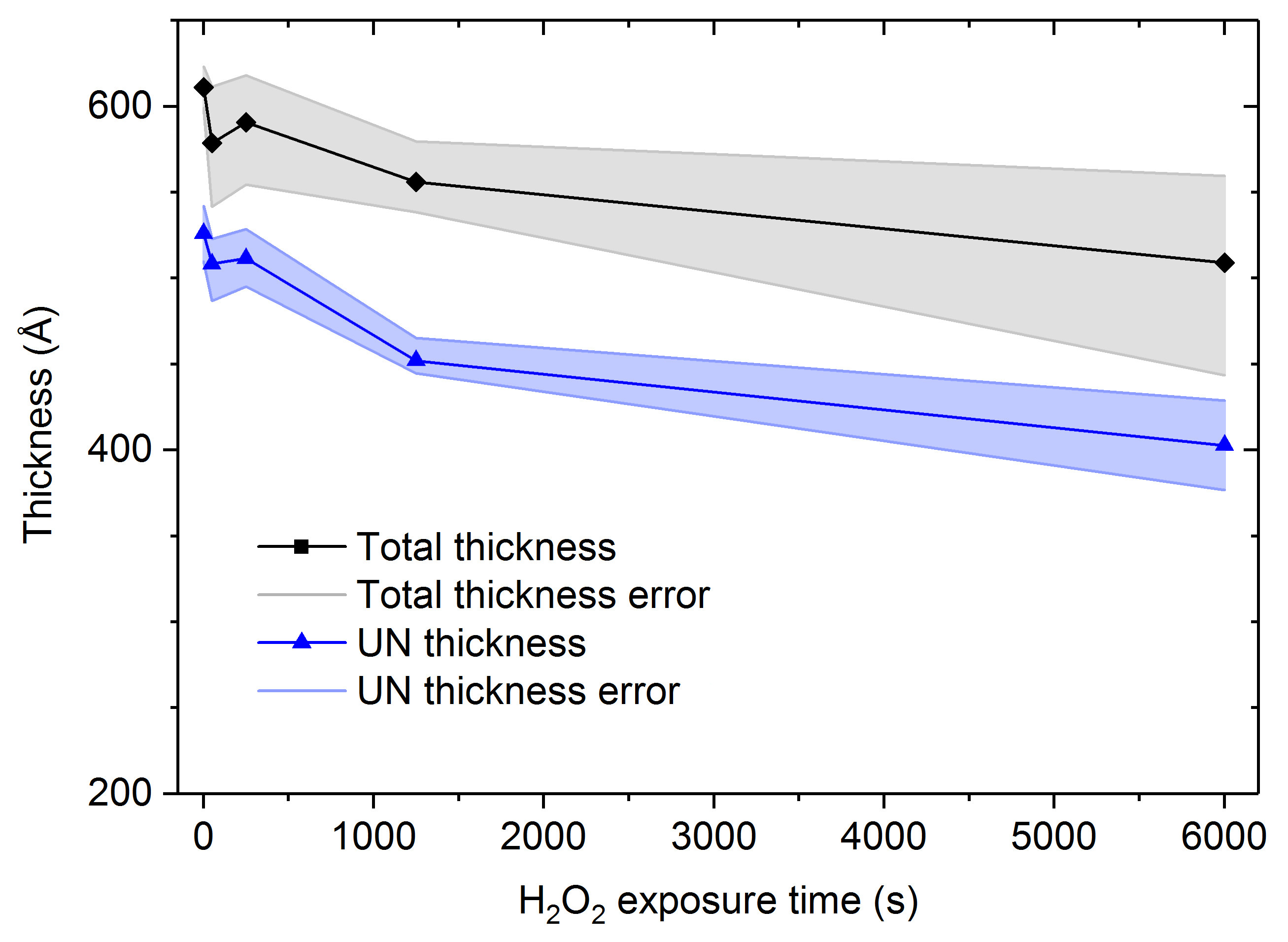}\caption{Total thickness (black) and UN layer thickness (blue) of the UN thin film samples as a function of H$_2$O$_2$ exposure time.\label{fig:unlayers}} \end{figure} 

Fig. \ref{fig:unlayers} gives a more detailed look at the results from the corrosion of the UN samples, showing change in UN film layer thickness in blue as well as total thickness in black.
The difference between the total thickness and UN layer thickness is composed of the UO$_2$ oxide layer and UO$_{2+x}$ higher oxide layer.
From this plot it can clearly be seen that the combined thickness of these layers is approximately constant as a function of H$_2$O$_2$ exposure, showing that the uranium oxide layer thickness after exposure to H$_2$O$_2$ is independent of exposure time.

\section{Discussion}

XRD analysis of UO$_2$, UN, and U$_2$N$_3$ thin films showed a single phase present in the case of UO$_2$ and U$_2$N$_3$.
The UN films were shown to contain some U$_2$N$_3$ contamination, however, given that the U$_2$N$_3$ phase is shown to corrode faster, this contamination is unlikely to play a role in the slow corrosion rate of UN.
Crystallite size analysis of the XRD data showed the samples to all have nanocrystalline grains of approximately 10\,nm for UN and U$_2$N$_3$ and 20\,nm for UO$_2$.

This experiment has shown that there are significant differences in material lost as a function of exposure time to H$_2$O$_2$, comparing UO$_2$, UN, and U$_2$N$_3$.
These materials were found to have corrosion rates of 0.083(3), 0.020(4), and 0.47(8)\,\AA{}/s, equivalent to 0.033(1), 0.010(2), and 0.19(3)\,mg/cm$^2$/hr, respectively, with a linear fit.
For a bulk material, exponentials or parabolics are often used to fit corrosion rates as surface area increases with breakaway oxidation or as the surface passivates, however, in the case of thin films, the surface remains smooth enough that surface area should not be a significant factor in corrosion rates \cite{Paljevic1975}.

The result that UN corrodes more slowly than UO$_2$ is surprising considering the literature on corrosion of UN in water, but also when it is considered that the surface of the UN sample is a UO$_2$ layer \cite{Sunder1998,Jolkkonen2017}.
XRR modeling showed the thickness of this UO$_2$ layer after exposure to H$_2$O$_2$ to be independent of exposure time.
As UN oxidises in air, it is not known if this oxide layer is of constant 90\,\AA{} thickness during exposure to H$_2$O$_2$ or if it decreases on exposure, re-forming when the film is removed and exposed to air.
This thickness corresponds to the passivating layer thickness consistently seen in these films.

While the UO$_2$ oxidised layer will be corroded on exposure, from the present data, it appears that this oxidised layer on the UN film corrodes more slowly than the UO$_2$ film as the exposure time required to corrode 90\,\AA{}, the oxidised layer thickness, is greater for UN than UO$_2$.
This suggests that there is a rate limiting step in the oxidation of UN to UO$_2^{2+}$ that is not present in the oxidation of UO$_2$ to UO$_2^{2+}$ or U$_2$N$_3$ to  UO$_2^{2+}$.
Rama Rao $et$ $al.$ state that the rate controlling process of oxidation of UN is the diffusion of nitrogen gas through the oxidised surface and out of the sample \cite{Rao1991}.
Were this to be the case, it would be expected that U$_2$N$_3$ would have a similarly slow corrosion rate, assuming micro-structural differences between the UN and U$_2$N$_3$ samples are not significant, but the results of this experiment show otherwise.

Not only are there differences in corrosion rates between the different materials, there are also differences is the way corrosion progresses.
UO$_2$ films showed a significant increase in sample roughness with exposure time, demonstrating that corrosion is not occurring uniformly over the film.
This can be seen in the roughness values from the modeled XRR, and is further supported by the appearance of a second critical angle after only 250\,s, corresponding to the glass substrate.
While the error on this data point is very large, it is still significantly higher than the roughnesses of both UN and U$_2$N$_3$.
This large difference is seen again at 1250\,s, despite the poor fit to the XRR data, where fringe depth in the data is lower than in the fit, suggesting roughness should be even higher than modeled.
This large increase in roughness of the UO$_2$ films as a function of H$_2$O$_2$ exposure time is possibly caused by relative differences in corrosion rates between grains and boundaries or different grain orientations, which have been shown to corrode at different rates for radiolytic dissolution \cite{Rennie2018}.

Conversely, the roughness of the U$_2$N$_3$ samples as a function of exposure time to H$_2$O$_2$ is unchanged within errors, showing that corrosion is progressing rapidly and uniformly across the film.
UN samples showed some increase in roughness with corrosion, but much lower than that of UO$_2$, suggesting that corrosion occurs fairly uniformly across the film.

Considering the significance of these results in the context of the nuclear fuel life-cycle, it could be argued that the 0.1\,M H$_2$O$_2$ concentration used is much higher than can be realistically expected to be caused by water radiolysis \cite{Ekeroth2006}.
However, this high concentration mimics the high corrosion rates that would be expected from other radiolytic products such as OH$^{\bullet}$ and H$^{\bullet}$ \cite{Springell2015} or in extreme scenarios such as during an accident. 
Additionaly, this high concentration reduces exposure times required to see significant differences in corrosion rates between materials, meaning that changes in H$_2$O$_2$ concentration due to decomposition does not introduce significant errors into the experiment.
While this experiment may not replicate the high temperatures and pressures that would be expected in an accident scenario, these initial tests indicate the need for further information to be aquired on the oxidation and corrosion of UN. 

\section{Conclusion}

XRR measurements showed different corrosion rates of UO$_2$, UN, and U$_2$N$_3$ thin films on exposure to 0.1\,M H$_2$O$_2$ for vaying times, with UN being the most corrosion resistant.
In this solution of simulated radiolytic products, corrosion rates were found to be 0.083(3), 0.020(4), and 0.47(8)\,\AA{}/s for UO$_2$, UN, and U$_2$N$_3$, respectively.
This result is in contrast with literature showing UN to corrode faster that UO$_2$ in water.
Data analysis also suggested a large increase in roughness as a function of H$_2$O$_2$ exposure time for the UO$_2$ films, which was not observed for the UN and U$_2$N$_3$ films.
These results show that UN could be more corrosion resistant in an accident scenario than previously thought, suggesting that it would be worthwhile continuing this investigation into radiolytically induced dissolution.
The observation of an oxide layer of consistent thickness on the UN film, independent of H$_2$O$_2$ exposure time, suggests that there is a  rate-limiting step in the oxidation of UN to UO$_2^{2+}$ that is not present in the oxidation of UO$_2$ or U$_2$N$_3$ to UO$_2^{2+}$ .
This is contradictory to literature on the oxidation mechanism of UN, highlighting the need for a better understanding of this process.

\section*{Acknowledgments}

The authors acknowledge funding from EPSRC grant 1652612.

\section*{Data availability}

The raw and processed data required to reproduce these findings will be made available on request.

\bibliography{Refs}{}

\end{document}